\documentclass[journal=ancac3]{achemso}
\setkeys{acs}{etalmode=truncate,maxauthors=20}
\usepackage[version=3]{mhchem} 
\usepackage[T1]{fontenc}       
\usepackage[utf8]{inputenc}
\usepackage[colorinlistoftodos,textsize=tiny]{todonotes} 

\author{Sven-Hendrik Lohmann}
\author{Christian Strelow}
\author{Alf Mews}
\author{Tobias Kipp}
\email{kipp@chemie.uni-hamburg.de}
\affiliation[Universität Hamburg]
{Institute of Physical Chemistry, University of Hamburg, Grindelallee 117, D-20146 Hamburg, Germany}

\title{Surface Charges on CdSe-Dot/CdS-Rod Nanocrystals: Measuring and Modeling the Diffusion of Exciton-Fluorescence Rates and Energies }

\begin{document}







\begin{abstract}
By performing spectroscopic single-particle measurements at cryogenic temperatures over the course of hours we study
both the spectral diffusion as well as the diffusion of the decay rates of the fluorescence emission of  core/shell CdSe/CdS dot/rod nanoparticles. A special analysis of the measurements allow for a correlation of data for single neutral excitons only, undisturbed by the possible emission of other excitonic complexes. We find a nearly linear dependency of the fluorescence decay rate on the emission energy. The experimental data is compared to self-consistent model calculations within the effective-mass approximation, in which migrating point charges set onto the surface of the nanoparticles have been assumed to cause the temporal changes of optical properties. These calculations reveal a nearly linear relationship between the squared electron-hole wave function overlap, which is linked to the experimentally determined fluorescence rate, and the exciton emission energy. Within our model single migrating surface charges are not sufficient to fully explain the measured rather broad ranges of emission rates and energies, while two -- and in particular negative -- surface charges close to the core of the DR induce large enough shifts. 
Importantly, for our nanoparticle system, the surface charges more strongly affect the hole wave function than the electron wave function and both wave functions are still localized within the dot-like core of the nanoparticle, showing that the type-I character of the band alignment between core and shell is preserved.
\end{abstract}

\vspace{2cm}
Semiconductor nanoparticles are discussed and already used for several kinds of applications, like in lighting or display devices. Nanoparticles as emitters benefit from their broad color tunability due to the strong size-dependence of their band gap\cite{norris_measurement_1996} and, furthermore, their ability to be processed in solutions.\cite{kim_contact_2008} 
The most severe problem for their use is the fluorescence intermittency of single particles\cite{neuhauser_correlation_2000}, which lessens the overall quantum yield and therefore the quality of such particles as emitters.\cite{ebenstein_fluorescence_2002} 
In light of this prevalently studied problem another phenomenon negatively influencing the emission properties of semiconductor nanoparticles is often overlooked. This is the spectral diffusion, which can be observed as random shifts of 
the emission wavelength over time.\cite{empedocles_photoluminescence_1996}

Spectral diffusion in spherical CdSe particles was first observed as an inhomogeneous line broadening at cryogenic temperatures, which was caused by energetic shifts within individual detection-time bins.\cite{empedocles_influence_1999} The energy shifts were of comparable strength as previously measured shifts induced by the quantum-confined Stark effect.\cite{empedocles_quantum-confined_1997}
Furthermore a correlation between the blinking behavior and spectral jumps was reported\cite{neuhauser_correlation_2000}. Spectral diffusion is explained by a change in the net surface charge, whereupon later studies suggest a discrete hopping of 
surface charges\cite{fernee_charge_2010}, which seems to be accompanied by a memory effect of the different charge positions\cite{fernee_spontaneous_2012}.

In the meantime core-shell structured nanoparticles were introduced that exhibited increased longterm spectral stability and decreased fluorescence intermittency.\cite{chilla_direct_2008} A special example of such a heterostructure particle consists of a spherical CdSe dot enclosed by a rod-like CdS shell\cite{talapin_highly_2003}, a so called dot-rod (DR). The shell itself, but especially the broken spherical symmetry give rise to improved optical properties such as improved optical gain\cite{grivas_single-mode_2013}, inherent linear polarization of the fluorescence emission,\cite{talapin_highly_2003,vezzoli_exciton_2015} as well as  possibilities to tune the optical parameters of the system. Accurate adjustments of the shell dimensions not only allow engineering the wave function overlap \cite{raino_probing_2011,granados_del_aguila_observation_2014}, but also tuning the splitting of the excitonic fine structure\cite{raino_controlling_2012,biadala_tuning_2014}. 
A possible quasi-type-II band alignment in these nanoparticles might allow for an easy manipulation of the electron and hole wave functions by external electric fields. In this case also migrating surface charges should have an increased influence on the emission.
Until now low temperature investigations on CdSe/CdS DRs have shown a correlation between the fluorescence peak intensity, peak width, phonon coupling strength and the fluorescence wavelength \cite{muller_monitoring_2004,muller_wave_2005}. The correlation has been qualitatively explained by a modification of the electron and hole wave functions by charges migrating on the DR surface. It has been argued that especially negative surface charges have a strong impact on the wave function distributions, in particular on the electron wave function, which was assumed to freely penetrate into the shell.\cite{muller_wave_2005} 

In this work we directly measured the correlation between spectral diffusion and changes in the fluorescence decay dynamics for CdSe/CdS DRs and compared it to model calculations, where point charges are positioned on the DR surface.
During the spectroscopic single-particle measurements at cryogenic temperatures over the course of hours,  consecutive spectra within time bins of 2 s and a full time-correlated single-photon counting (TCSPC) data set in the so-called time-tagged time-resolved (TTTR) mode were recorded simultaneously. These combined datasets allow for an analysis of the emission properties of single neutral excitons only, isolated from the emission of other (charged) excitonic complexes. 
The correlation of the energy- and time-resolved data reveals a monotonic dependency of the fluorescence decay rate on the emission energy of the single neutral exciton, which is shifting over time (spectral diffusion). 
To link the spectroscopic observations to a quantitative picture, we performed model calculations within the effective-mass approximation in which migrating surface charges are described by point charges modifying the potential environment rather than by homogeneous electric field gradients as used in simulations of the quantum-confined Stark effect. 
The calculations revealed that the spectral diffusion as well as the accompanying change in the recombination dynamics can in principle be assigned to point charges migrating on the DR surface. We find that charges close to the dot region have the strongest impact, but single elementary positive or negative point charges are not sufficient to explain the rather wide range of energy shifts and lifetime changes. A good agreement between experiment and simulation is found for configurations of two negative surface charges. In contrast to common assumptions, we can show that for our DR system the surface charges have a stronger effect on the hole, which is confined in the core, than on the electron, which is often assumed to expand into the rod. The electron is still localized in the DR core, which demonstrates the preservation of the  type-I character of the band alignment in our DR system.

\section{Results/Discussion}
\begin{figure}[t]
  \includegraphics{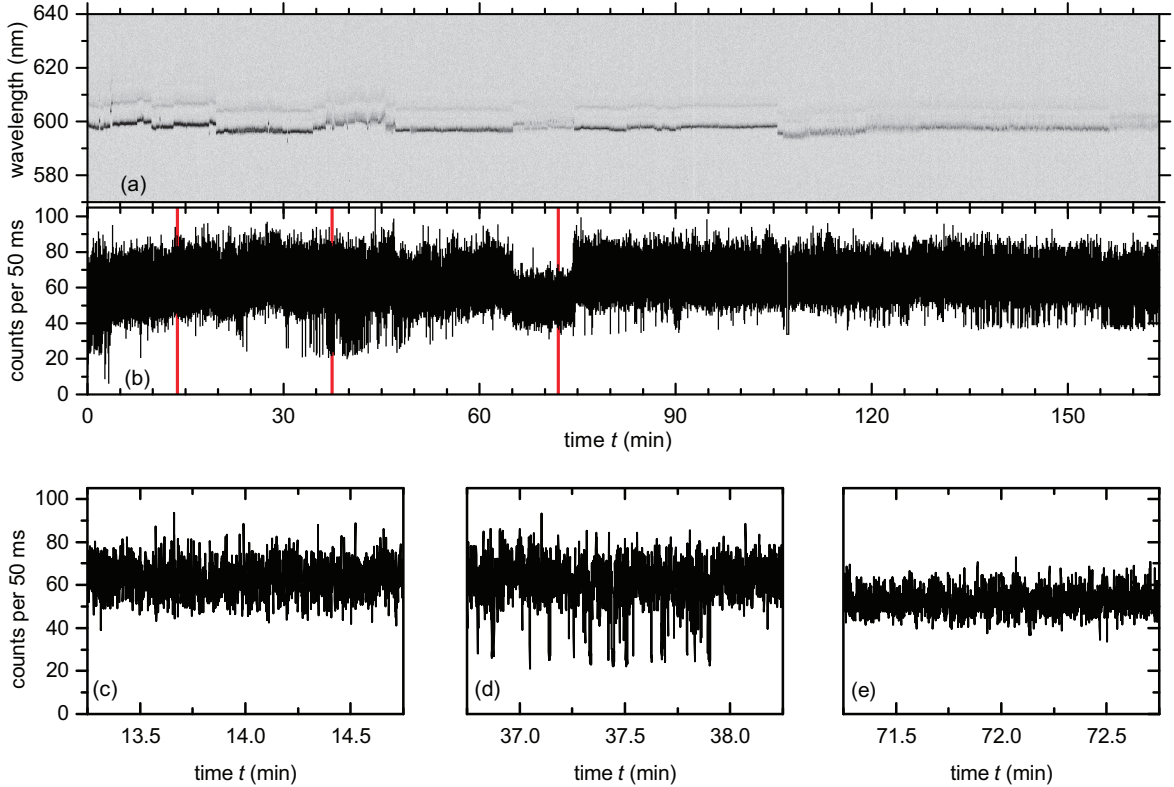}
    \caption{(a) Time trace over 2 h 45 min of subsequently recored PL spectra, each obtained with a CCD camera integration time of 2 s. The PL intensity is coded as a gray scale where dark means high intensity. (b) Time trace of the integrated PL intensity for a time bin set to 50 ms. The trace has been derived from TTTR data collected with an APD. (c-e) Segments of the intensity time trace for the three 90-s-time periods marked by red bars in (b).}
  \label{figure1}
\end{figure}

Figure \ref{figure1} (a) shows the temporal evolution of the emission spectra (each recorded within 2 s integration time) of an individual DR over the course of 2 hours and 45 minutes. The spectra are dominated by one peak, the so-called zero-phonon line (ZPL), which occurs at a wavelength slightly below 600 nm for time $t=0$\,min, but which spectrally shifts and jumps over the course of the measurement. Red-shifted to the ZPL, an additional satellite peak with lower intensity occurs. The constant distance to the ZPL identifies this peak as the first replica of the LO phonon of the CdSe core material.
Figure \ref{figure1} (b) shows the intensity time trace of the DRs obtained from the TTTR data with the time bin set to 50 ms. It has been corrected for a slow and monotonous decrease of the detected emission over the course of the experiment, which can be seen in Fig.\ \ref{figure1} (a) and which we attribute to a small thermal drift of the focal point of the objective with respect to the sample. The intensity time trace reveals an extraordinary stability in the emission of the DR. In particular, over the whole 165 min of measurement the emission never went completely off within the 50 ms time bins. Instead one can identify roughly three different emission levels. The most intense level (close to 60-70 counts per 50 ms) is also the most frequent level over the course of the whole measurement. Figure \ref{figure1} (c) shows an exemplary segment of the time trace for a scaled-up period of 90~s around $t=14$\, min, in which solely this highest intensity level occurs. In contrast to this, for example for $30 ~\textrm{min} < t < 50~\textrm{min}$, but also in other periods of time, a smaller intensity level (roughly between 20 and 40 counts per 50 ms) occurs, very rapidly alternating with the high intensity level. This behavior is represented in the exemplary 90-seconds segment of the time trace around $t=37.5$\, min shown in Fig.\ \ref{figure1} (d). Interestingly, a third intermediate intensity level occurs, which is for example exceptionally prominent and stable for $65~\textrm{min}<t<75~\textrm{min}$. A corresponding 90-seconds segment of the time trace is shown in Fig.\ \ref{figure1} (e). 
We attribute the different intensity levels to different excitonic complexes inside the DR. In particular, we assign the dominant high-intensity level to the radiative recombination of the simple uncharged single exciton. The other levels might be assigned to charged excitons. A detailed analysis of the different emission levels, the corresponding emission energies and fluorescence lifetime is not within the scope of this work. Here we want to solely concentrate on the effect of spectral diffusion that occurs in the temporal evolution of the DR spectra. For our analysis, we concentrate only on the dominant single neutral exciton emission. To select data that is unaffected by the presence of other excitonic complexes, we categorized every 50-ms-time bin of the time-trace  by being either above or below an intensity threshold, thus corresponding either to the single neutral exciton emission or not. In a next step, we assign only those of the 2-s-time bin spectra to "pure" single neutral exciton emission, if at least 35 of the 40 time bins of the time traces have been assigned to the single exciton case. By this procedure, we identified 1588 of the 4950 individual spectra shown in Fig.\ \ref{figure1} (a) as originating from single neutral exciton emission. Consequently, all other spectra have been discarded for the following analysis.

\begin{figure}[t]
  \includegraphics{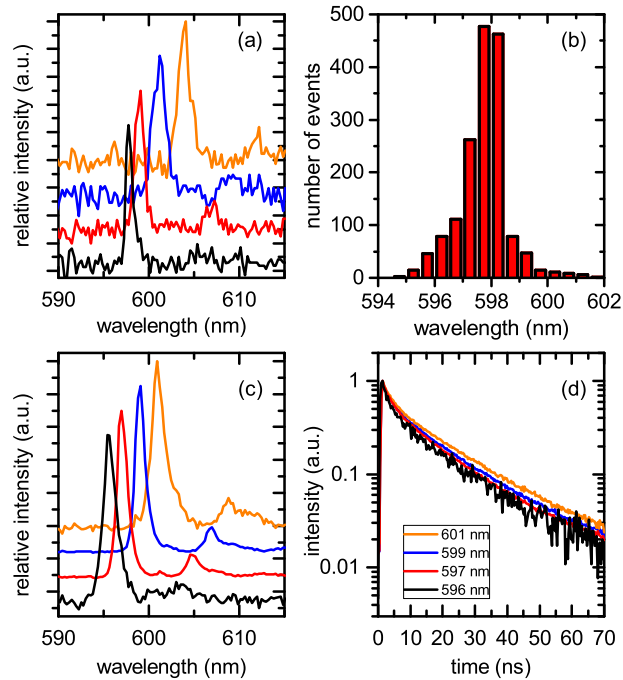}
  \caption{(a) Exemplary single-exciton spectra recorded within 2 s integration time. The spectra are normalized to their maximum and vertically shifted for clarity. (b) Histogram of the single-exciton zero-phonon-line emission wavelengths. (c) Exemplary averaged single-exciton spectra, normalized and vertically shifted for clarity. (d) Fluorescence decay curves corresponding to the averaged spectra shown in (c).}
  \label{figure2}
\end{figure}

Figure \ref{figure2} (a) shows four exemplary spectra, each with a single-exciton ZPL at different wavelengths. Red-shifted to the ZPLs, faint phonon side bands can be recognized, as already discussed with respect to Fig.\ \ref{figure1} (a).
For an in-depth analysis of the emission-wavelength dependency of particularly the fluorescence-decay dynamics but also of the ZPL width and the exciton-phonon coupling strength, we fitted the ZPL of each spectrum by a Lorentzian and used the wavelength positions of the maxima as a criterion to categorize the spectra into wavelength intervals of 0.5\,nm.
The resulting histogram of the emission wavelength, depicted in Fig.\ 
\ref{figure2} (b),  reveals a distribution in the range between 594 and 602~nm. 
Figure \ref{figure2} (c) shows averages of such categorized spectra for different wavelength. Due to an increased signal-to-noise ratio the phonon side bands are now more clearly visible and can be assigned to LO phonon replica of the \ce{CdSe}-core material with an energy distance of 25~meV\cite{chilla_direct_2008} and to the corresponding replica of the \ce{CdS}-shell material with an energy distance of 35~meV,\cite{granados_del_aguila_observation_2014} which appears as a less pronounced shoulder in the spectra.
The simultaneous and synchronized acquisition of spectrally-resolved data and TTTR single-photon counting data allows for the construction of decay curves that exactly correspond the averaged spectra. Figure \ref{figure2} (d) shows such decay curves for the spectra depicted in panel (c) on a semi-logarithmic scale. 
All such decay curves can be well approximated by bi-exponential functions with  a fast lifetime component around 2 ns and a longer lifetime component between 13 and 20 ns. 

After this data processing it is now possible to analyze the spectral diffusion of the single exciton emission in our DRs in depth. Here, the correlation of the spectral diffusion to the fluorescence decay rate is of special importance since, as we will elucidate later, it allows for a semi-quantitative modeling of the spectral diffusion and its accompanying changes in the exciton decay dynamics induced by migrating surface charges. 

\begin{figure}[t]
  \includegraphics{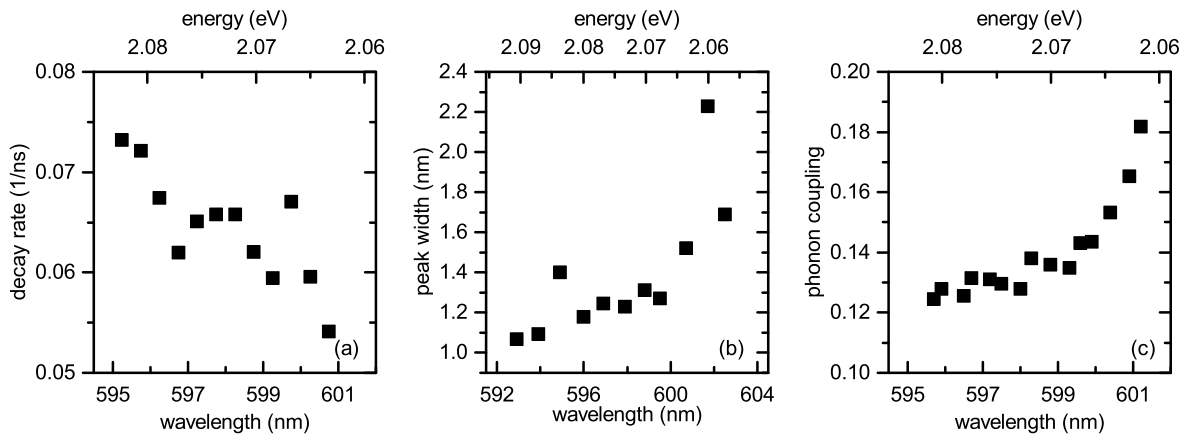}
  \caption{Wavelength dependency of (a) the decay rate, (b) the peak width, and (c) the phonon coupling of the spectrally diffusing single exciton emission.} 
  \label{figure3}
\end{figure}

The excitonic level structure in DRs has been extensively studied and it is commonly accepted that, above the vacuum ground state, the first and the second single-exciton states are separated by $\Delta E$ because of the electron-hole exchange interaction.\cite{raino_controlling_2012,biadala_tuning_2014} The state higher (lower) in energy is the so-called bright (dark) state for which the optical transition into the ground state is allowed (forbidden). For low temperatures $(k_B T \ll \Delta E)$, lifetime measurements reveal a biexponential decay with the short component in the range of 1 ns essentially representing the spin-flip rate from bright to dark exciton state and the long component in the 100-ns range representing the radiative recombination rate of the nominally forbidden transition \cite{biadala_tuning_2014}. With increasing temperature, the bright state is thermally populated, leading to a decrease of the intensity of the short component and a shortening of the long component, which then represents a mixture of the radiative recombination rates of the bright and the dark exciton states.\cite{biadala_tuning_2014} Since these two rates should be directly affected by the electron-hole wave function overlap, we evaluated the longer lifetime component of the decay curves constructed for each wavelength interval of the histogram in Fig.\ \ref{figure2} (b). Figure \ref{figure3} (a) shows the respective decay rate \latin{versus} the corresponding emission wavelength.  It reveals a decrease of the decay rate with increasing emission wavelength. 

Before we analyze this behavior in detail and compare the experimental result to a theoretical model in order to get inside into the mechanism of spectral diffusion, we want to note that our data processing also allows for a correlation of the spectral diffusion of the single-exciton emission to the ZPL peak width and to the phonon-coupling strength, as is shown in Fig. \ref{figure3} (b) and (c), respectively. 
Both the single-exciton ZPL peak width and the single-excition phonon coupling strength are increasing with increasing emission wavelength. Both trends have been previously observed by Müller \latin{et al.}\cite{muller_wave_2005} and were qualitatively explained by surface charges dominantly manipulating the electron wave function localization in a DR system. The underlying assumption for this explanation is that in that DR system, the hole is confined in the CdSe core, whereas the electron is free to penetrate the CdS shell, \latin{i.e.}, the DRs essentially exhibit a quasi-type-II band alignment. 
By scanning tunneling spectroscopy on DRs, the band offsets between CdSe and CdS could be directly measured. Both materials were found to form type-I heterostructures with a conduction band offset of about 0.3 eV,\cite{steiner_determination_2008} in line also with previous theoretical calculations \cite{wei_first-principles_2000}. In CdSe/CdS DRs, the size of the core in particular determines the electron confinement. For decreasing core diameters, a transfer from a type-I localization to a quasi-type-II delocalization has been reported for core diameters of about 2.8 nm.\cite{sitt_multiexciton_2009}. Since our dot-in rods have much larger core diameter of about 4.7 nm we can safely assume that here a type-I localization of electrons and holes occur. 

To link our spectroscopic results, in particular the correlation of emission wavelength and decay rate from Fig.\ \ref{figure3} (a), to a quantitative microscopic picture of the charge carriers involved in the emission process, we performed model calculations. We described our DR system within the two-band effective-mass approximation by self-consistently solving the Schrödinger equation for electrons and holes on a three-dimensional spatial grid. 
The potential terms consist of the external potential for electrons or holes, given by the conduction- and valence-band offsets, respectively, and the iteratively obtained Coulomb potential induced by the opposite carrier. We assumed a type-I band alignment with offsets of 300 meV and 380 meV for conduction and valence band, respectively. The CdSe bulk bandgap was assumed to be 1.74 eV. For the particle geometry, we set the DR length to 20 nm, the width to 5.2 nm and the core diameter to 4.7 nm, as obtained by averaging values obtained by transmission electron microscopy. We modeled the effect of additional charges localized at the DR surface by modifying the external potential with respect to the Coulomb potential of the surface charges. For the effective electron and hole masses in CdSe and CdS, we assumed $m_\textrm{e}^\textrm{CdSe}=0.13$, $m_\textrm{h}^\textrm{CdSe}=0.45$, $m_\textrm{e}^\textrm{CdS}=0.20$, and $m_\textrm{h}^\textrm{CdS}=0.70$ \cite{mews_preparation_1994,haus_quantum_1993}, respectively. The relative permittivity inside the DR was set to 9.5, \cite{madelung_cadmium_nodate} while outside a value of 2.0 was assumed, which should roughly emulate the situation for ligands (\latin{e.g.}, trioctylphosphine oxide with a permittivity of 2.6 \cite{jiang_electrostatic_2000}) rather densely attached to the DR surface. Image charges because of the dielectric mismatch have not been taken into account. The calculations reveal electron and hole wave functions and exciton energies for different surface charge configurations.

 \begin{figure}
  \includegraphics{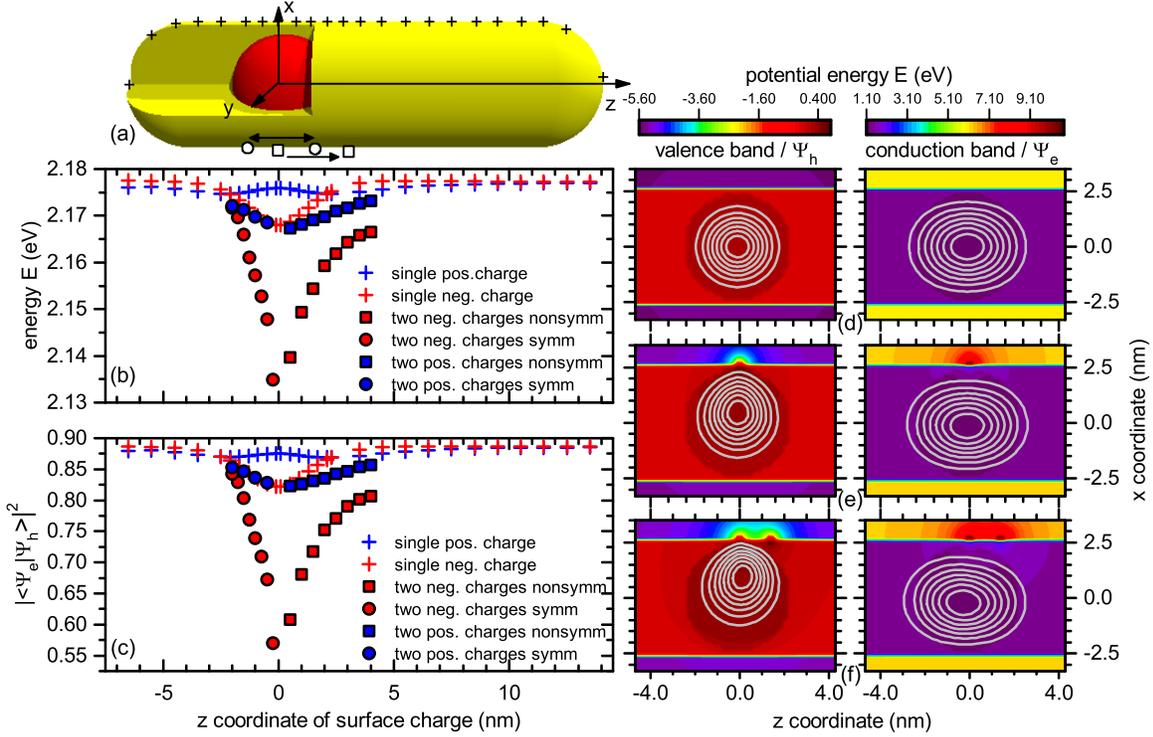}
  \caption{(a) Sketch of a DR. The origin of the coordinate system is places in the center of the dot. Crosses represent positions of single elemental surface charges used in calculations. Squares sketch the configuration of two surface charges, one fixed at $z=0$ and the other at variable $z$ position. Circles represent two surface charges with variable positions $-z$ and $z$. (b) Calculated exciton energy $E$ for one (crosses) or two (circles and squares) positive (blue) or negative (red) elemental point charges on the DR surface. The positions of the charges are given by the variable $z$ (crosses), by $-z$ and $z$ (circles), and by $z=0$ and the variable $z$ (squares). (c) Calculated squared wave function overlap integrals $| \langle \Psi \textsubscript{e} | \Psi \textsubscript{h} \rangle |^2$ for the different surface charge configurations. (d-f) Two-dimensional color plots of the self-consistent valence (left) and conduction (right) band potentials of the DR in the $xy$-plane (d) without surface charges, (e) with one negative charge at $z=0$, and (f) two negative charges at $z=0$~nm and $z=1.4$~nm.  In each plot, the white contour lines represent the corresponding calculated self-consistent hole or electron ground state wave function  ($\Psi_\textrm{h}$ or $\Psi_\textrm{e}$).}
  \label{figure4}
\end{figure}

Figure \ref{figure4} (a) sketches a DR, the coordinate system we refer to in the following, and exemplary positions of the assumed surface charges. 
Figure \ref{figure4} (b) shows the calculated exciton energies $E$ in dependence of $z$, where $z$ parameterizes positions of surface charges as will be explained in the following. Crosses in Fig.\ \ref{figure4} (b) represent individual surface charges at the positions $z$. It can be seen that only surface charges in the direct vicinity of the core ($z \approx 0$) have a perceptible impact on the exciton energy. Individual positive elemental surface charges (blue crosses) can modify the energy by not more than 3 meV. On the other hand, a single negative elemental charge can decrease the energy by up to 10 meV, as can be inferred from the red crosses in Fig.\ \ref{figure4} (b), indicating a minimum energy for $z=0$. Since in our experiment we observed spectral shiftings up to 20 meV, we also calculated the influence of two elemental surface charges on the exciton energy, representing the simplest situation beyond single charges. Exemplary, blue-filled squares in Fig.\ \ref{figure4} (b) represent exciton energies for two positive surface charges, one set to $z=0$ and the other set to variable $z$. Blue-filled circles represent the situation of two positive surface charges placed symmetrically around $z=0$ at variable positions $-z$ and $z$. Both of these configurations yield maximum energy shifts in the same range as a single negative surface charge. The red-filled squares and circles in Fig.\ \ref{figure4} (b) represent the same geometric surface-charge configuration as the blue-filled counterparts but for negative elemental charges. It can be seen that here, considerably larger shifts can be expected, easily exceeding the measured shifts of about 20 meV. Principally there is a wealth of other possibilities to place two charges on the surface of a DR. In order to have a considerable effect on the exciton energy, however, both charges have to be placed in the vicinity of the core. This point will be addressed later, again. Comparing the different impact of positive and negative charges, the calculations reveal that the steep potential 
of positive surface charges close to $z=0$ leads to a decrease of the electron eigenenergy that is on the other hand nearly compensated by the repulsion, \latin{i.e.} a further confinement, of the hole.
Negative surface charges around $z=0$ repel the electron wave function and by that increase their energies, but their impact on the hole is stronger. Due to its larger effective mass the hole binds stronger to the negative surface charge, thus the increase of the electron eigenenergy is overcompensated by the decreased hole eigenenergy. This explains the stronger impact of negative surface charges than positive ones.
Note, that the changed confinement energy is not only influenced by the effective masses but also by the external confinement potential. Since in our case the system gives rise to a type-I band alignment we find a rather strong confinement potential for both electron and hole. In case of a more flat band alignment (or even a type-II band alignment) of the conduction band the electron wave function would be influenced much stronger by the repulsion of a negative surface charge and thus the properties of the spectral diffusion and the diffusion of the decay rate might be governed by the electron.

Figures \ref{figure4} (d-f) visualize the self-consistently calculated valence band (left) and conduction band (right) potential landscape in the $xz$-plane for exemplary surface charge configurations in a false color scale, together with the calculated hole and electron ground state wave functions as white contour lines. 
For no surface charge (panel d) the potentials as well as the wave functions are rotationally symmetric with respect to the $z$-axis; the electron and hole wave functions are localized in the center of the DR's core.  Consequently the overlap of electron and hole wave function is largest for this case. 
For a single negative charge at $z=0$ (panel e) and, much more pronounced, for two negative charges non-symmetrically positioned at $z=0$~nm and  $z=1.4$~nm (panel f), the hole wave function is attracted in direction of the charge. The electron wave function is slightly rejected by the charges but, importantly, the electron is still clearly localized in the core of the DR. As a consequence of the wave function redistributions the squared overlap integral $| \langle \Psi \textsubscript{e} | \Psi \textsubscript{h} \rangle |^2$ is decreased in these cases of surface charges.
Figure \ref{figure4} (c) depicts calculated values of $| \langle \Psi \textsubscript{e} | \Psi \textsubscript{h} \rangle |^2$ \latin{versus} the $z$ position of the different surface-charge configurations (as defined in above discussion of Fig.\ \ref{figure4} (b)). Obviously the overlap exhibits a very similar behavior as the exciton energy shown in Fig.\ \ref{figure4} (b). 

 \begin{figure}
  \includegraphics{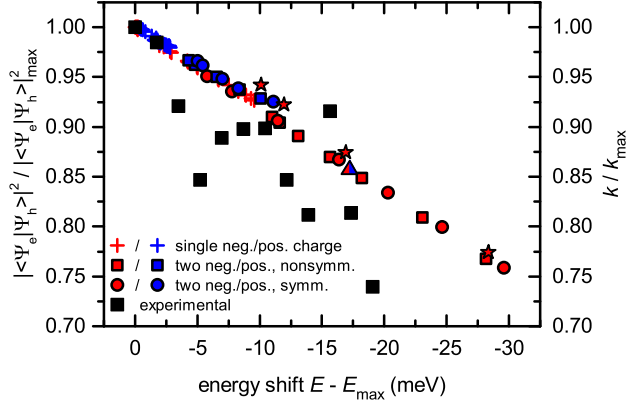}
  \caption{Calculated values of the normalized squared wave function overlap $| \langle \Psi \textsubscript{e} | \Psi \textsubscript{h} \rangle |^2 / \langle \Psi \textsubscript{e} | \Psi \textsubscript{h} \rangle |^2_\textrm{max} $ vs.\  the calculated energy shift $E-E_\textrm{max}$ for the different surface-charge configurations (red and blue crosses, squares, and circles similar as in Fig.\ \ref{figure4} (b,c); stars and triangle defined in the text) compared to experimentally determined data pairs of normalized decay rates $k/k_\textrm{max}$ and corresponding energy shift (black squares).}
  \label{figure5}
\end{figure}
Combining both panels b and c of Fig.\ \ref{figure4} yields that a pronounced shift in energy comes along with a considerable decrease of the overlap integral. This behavior is summarized in Fig.\ \ref{figure5} where the normalized squared wave function overlap $| \langle \Psi \textsubscript{e} | \Psi \textsubscript{h} \rangle |^2 / \langle \Psi \textsubscript{e} | \Psi \textsubscript{h} \rangle |^2_\textrm{max} $ is plotted against the energy shift $E-E_\textrm{max}$ for the different surface-charge configurations, using the same data-point designs as in Fig.\ \ref{figure4}. The data points for a single negative charge cluster at small values for the energy shift. 
The largest spread in terms of energy shift and wave function overlap for the surface-charge configurations considered so far occurs for two negative charges.
All the different surface-charge configurations yield a nearly linear decrease of the squared overlap with decreasing exciton energy with only slightly different slopes.
According to Fermi's golden rule, the squared wave function overlap of electron and hole is proportional to their radiative recombination rate. Hence, the calculated values should be connected to the experimentally obtained decay rates for different emission wavelengths. The black squares in Fig.\ \ref{figure5} represent the measured decay rates as already shown in Fig.\ \ref{figure3} (a). 
The difference here is that we normalized the decay rates to their maximum ($k/k_\textrm{max}$) and examine the energy shifts with respect to the maximum emission energy in order to plot these values on the same scales as the calculated ones. 
Doing so, we inherently assume that the largest measured decay rate corresponds to the largest calculated electron-hole wave function overlap.

The comparison reveals that the experimental data exhibits an energy-shift dependency with a similar slope as the calculated data sets. 
It becomes obvious that single charges, either positive or negative, are not sufficient to explain the measured large ranges of energy shifts or different decay rates. Increasing the complexity and assuming two surface charges yields that in particular two negative surface charges, depending on their exact positions, can cover the observed ranges.
In general, the closer the surface charges are the stronger are the resulting decreases in exciton energy and electron-hole wave-function overlap. However, it should be noted that the mutual approach of migrating surface charges to very small distances becomes unlikely because of their electrostatic repulsion. 

In our calculations so far, we assumed the surface charges to be aligned in axial direction.
Of course, as already mentioned, in principle every other configuration might also occur. In the following we systematically consider these other configurations. Generally, for all other configurations it still holds that charges close to the core of the DR have the strongest impact on the exciton energy and wavefunction. Consequently, two negative surface charges azimuthally aligned with $z=0$ represent an antipole configuration with respect to the axially aligned charges. All other possible configurations of two surface charges can be regarded as partly axial and partly azimuthal and their impact on the exciton energy and wavefunction should be smaller than corresponding axially or azimuthally aligned configurations. We modeled the azimuthal configuration of two negative surface charges at $z = 0$, with their position vectors forming angles of 180$^\circ$, 135$^\circ$, 90$^\circ$, and 45$^\circ$. Corresponding data points for the normalized wave function overlap and the energy shift are shown in Fig.\ \ref{figure5} as red stars. They all follow a similar trend as the other calculated points. Two opposing charges (180$^\circ$) lead to an energy shift of about 10 meV and a slightly larger wave function overlap than other surface charge configurations with similar energy shifts. For decreasing azimuthal angles the data points naturally approach the data points calculated for two negative surface charges aligned in z-direction (red squares and circles). 
Until now, only charges of the same sign have been regarded. One might also think about two charges of opposite sign impacting on the exciton. This impact should be maximal for both charges opposing each other at $z=0$; the result of the corresponding calculation is represented by the red-blue triangle in Fig.\ \ref{figure5}. Interestingly, this data point with its energy shift of about 17 meV also fully lies within the trend defined by all other surface-charge configurations. 
We note that, in principle, not only one or two but even more surface charges could influence the exciton in the DRs. The number of possible configurations, however, makes a modeling very complex, while the physical meaningfulness of such calculations, in which opposite charge carrier can partly cancel themselves, is limited. Importantly, the calculations reveal that the energy shift induced by only one surface charge is too small while two surface charges are enough to explain the range of measured shifts. We find a nearly linear correlation between the squared wave function overlap integral and the energy shift induced by surface charges. This calculated linear behavior is linked in a reasonable way to the monotonous dependency of the normalized decay rates on the shift of the emission energy.

Very recently, Sercel \latin{et al.}\ published a theory paper on photoluminescence enhancement through symmetry breaking induced by defects in nanocrystals. \cite{sercel_photoluminescence_2017} Using a pertubative approach the authors calculated that Coulomb centers inside small spherical core-only CdSe NCs can significantly break the selection rules responsible for the dark excitons and lead to an exchange of oscillator strength between fine structure states with different Bloch wavefunctions. A superposition of emission from all excitonic fine structure levels in thermal equilibrium then gives theoretical values for the experimentally accessible fluorescence lifetime. 
According Sercel \latin{et al.}\ the level mixing and the exchange of oscillator strength is influenced by a complex interplay between nanoparticle size and geometry and position of the Coulomb center. Although these effects become weaker for larger nanocrystals and for charges on the surface, it stays an interesting question how these findings can be extrapolated to large asymmetric core/shell particles with surface charges like in our system. 
Since the mixing of the Bloch wavefunctions is not included in our model, future work has to clarify if the level mixing in the case of our system can improve the explanation of the experimental results beyond our already good description by a change of the envelope wavefunction.

Finally, we want to discuss the possible origin of surface charges provoking the spectral diffusion. In essentially uncharged DRs, surface trap states might get populated with carriers of previously photoexcited electron-hole pairs. However, trapping of either the electron or the hole alone would lead to the creation of trions, \latin{i.e.}, charged excitons, inside the dot-in-rods during the next photoexciation process. These trions exhibit different recombination dynamics because of the Auger mechanism. Our in-depth analysis of spectrally-resolved emission time traces ensured that we are only dealing with the spectral shifting of the single neutral exciton emission. Consequently, one might speculat that both the photogenerated electron and hole get trapped on the surface. These surface charges then influence the emission energy of the consecutively photogenerated electron-hole pairs. Spectral shifting in time will then occur, e.g., upon trapping of further electron-hole pairs or upon hopping of surface charges between different surface-trap sites. It has also been speculated that charges are fluctuating within the ligand layer.\cite{braam_role_2013} Instead of originating from photoexcitation, one might also think about surface charges as an effect of a detachment and attachment of ligands surrounding the nanocrystals. A detachment of phosphonate ligands (from octadecyl or hexyl phosphonic acid, as used in our synthesis), which act as X-type ligands \cite{boles_surface_2016}, might induce positively charged surface sites. An extra ligand would induce a negatively charged surface site. Spectral shifting over time will then occur by spatial diffusion of mobile ligands, as has been rationalized in recent DFT calculations \cite{voznyy_mobile_2011}. Ultimately, the exact origin of the fluctuating surface charges is not well understood and further experimental and theoretical work has to be done to unravel the effects of surface chemistry and physics for nanocrystals with their different facets and differently coordinated surface atoms.

\section*{Conclusions}
In conclusion, we measured the temporal evolution of the photoluminescence emission of individual CdSe/CdS DRs at low temperature over the course of hours. We simultaneously recorded a sequence of emission spectra with integration-time bins of 2 s and a synchronized set of time-tagged time-resolved data by time-correlated single-photon counting. 
The emission spectra are dominated by the ZPL of exciton recombination, revealing some spectral shifts and jumps. The spectrally integrated fluorescence time trace constructed from the TTTR data reveals about three different intensity levels, of which the high-intensity emission level is most frequent. We attribute this emission to the recombination of single neutral excitons. For the further analysis of the spectral diffusion, only spectra originating from single neutral excitons have been evaluated. First, single-exciton spectra have been sorted by their peak wavelength and spectra with the same peak wavelength have been averaged. The synchronized spectral- and time-resolved measurement then allowed for the construction of decay curves that are exactly associated to the corresponding averaged spectra. The analysis of the decay rates revealed a decrease for increasing emission-peak wavelengths. In order to explain this behavior and to get insight into the physical reason of the spectral diffusion, self-consistent model calculations within the effective-mass approximation have been performed. We calculated the exciton energy as well as the electron and hole wave function overlap within the DR for different configurations of surface charges on the DR.
We find that surface charges only affect the exciton significantly, when placed close to the core of the DR. 
Then, the surface charges generally decrease both, the exciton energy as well as the overlap integral, while negative surface charges have a stronger impact than positive ones.
Unlike it is often assumed, in our DR system the type-I character of the band alignment is still preserved. Thus it is not the electron that is affected most by the surface charges, even though its wave function can more easily penetrate into the shell. Instead, it is the hole with its larger effective mass that can effectively be bound to a negative surface charge, by which the exciton energy and the overlap integral is strongly changed.
Analyzing different surface-charge configurations, the normalized squared electron-hole wave function overlap decreases basically linear with the decreasing exciton energy. 
Comparing these calculated values with the measured data pairs of the normalized decay rate and the emission-energy shift yields a good congruence.
The calculations reveal that single migrating surface charges are not sufficient to cause the measured rather broad ranges of decay rates and emission energies, while two surface charges close to the core of the DR lead to large enough shifts to explain the measurements. 
It is important to note that for configurations of one or two surface charges, both the electron and the hole wave functions are always localized within the core of the DR. This means that in our system, there is no quasi-type-II band alignment, which would make it possible for the electron to essentially spread over the whole rod-like shell surrounding  the dot-like core of the nanoparticle. 
Such a type-II kind of behavior, which would lead to a considerably stronger susceptibility of the optical properties on migrating surface charges but also on external electric fields, might only occur in DRs with considerably smaller cores\cite{sitt_multiexciton_2009} or probably in DRs with specifically designed lattice-strain distributions\cite{jing_insight_2015}.

\section{Methods/Experimental}
The CdSe/CdS DRs were fabricated following the synthesis by Carbone and coworkers.\cite{carbone_synthesis_2007} The DRs exhibited a core diameter of about 4.7 nm and an overall width of about 5.2 nm and length of about 20 nm, as determined by transmission-electron microscopy (TEM).
For single-particle measurements the DRs were diluted in hexane and spin-coated onto a silicon substrate. The sample was mounted into a closed-cycle cryostat (Attocube AG, temperature $\sim$5 K) as a part of a home-built confocal laser scanning microscopy setup. For excitation a 514 nm pulsed diode laser (Picoquant, 10~MHz pulse rate, \textless~70~ps pulse width) was used. A microscope objective (100x, 0.8 NA) both focused the excitation light and collected the emission. A beam splitter was placed in the detection path guiding equal parts of the light simultaneously to (i) a grating spectrometer equipped with CCD camera and (ii) to an avalanche photodiode (APD) for time-tagged time-resolved (TTTR) measurements using the reverse start-stop mode (PicoHarp 300, PicoQuant GmbH), allowing for simultaneous acquisition of (i) spectrally and (ii) time-resolved data.

The self-consistent calculations of electron and hole wavefunctions and eigenenergies in DRs with and without surface charges rely on a three-dimensional generalization of a one-dimensional approach to solve a Schrödinger equation with an arbitrarily shaped potential, as was described in detail in Ref. \cite{strelow_light_2012}.
We performed similar three-dimensional calculations for exciton complexes in CdSe quantum nanowires. \cite{franz_quantum-confined_2014}. They rely on the discretization of the single-particle Schrödinger equation on a
three-dimensional spatial Cartesian grid and a diagonalization of the emerging system of linear equations. The nearest distance between adjacent grid points was set to 2.0 \r{A} in all three dimensions. The number of grid points was  $35 \times 35 \times 110$.

\begin{acknowledgement}

S.-H.~L. thanks the Fonds der Chemischen Industrie (FCI) for financial support.
T.~K. acknowledges funding from the European Union's Horizon 2020 research and innovation programme under the Marie Sk{\l}odowska-Curie grant agreement No. 656598.

\end{acknowledgement}


%


\bibliography{PaperSpektraleDiffusion}

\newpage
\section{Graphical TOC Entry}
\begin{figure}
  \includegraphics{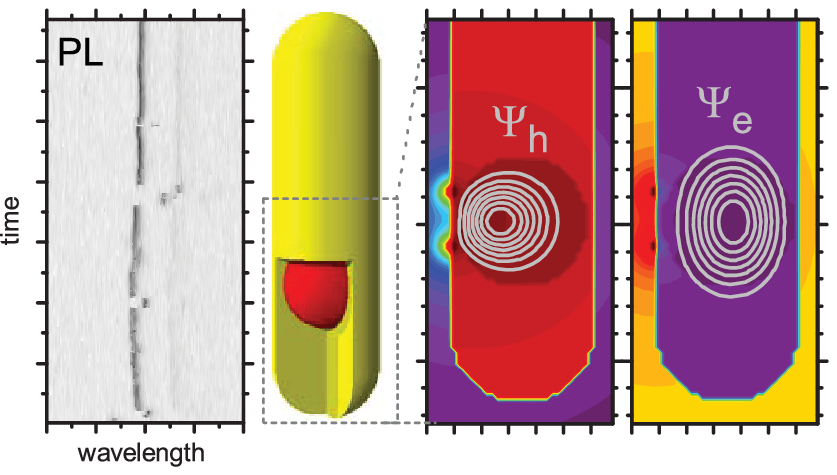}
\end{figure}

\end{document}